\documentclass[letterpaper,twocolumn,appendixfloats]{emulateaj}
\usepackage{float}
\usepackage{epsfig}
\usepackage{graphicx}
\usepackage{graphics}
\usepackage[latin1]{inputenc}
\usepackage{latexsym}
\newcommand{\nd}{\multicolumn{1}{c}{$\dots$}}

\newcommand{\nbrt}{20{,}000}
\newcommand{\nbrs}{9{,}125}
\newcommand{\nvar}{188}
\newcommand{\nnew}{67}
\newcommand{\nold}{121}
\shorttitle{Photometry from Dome A}
\shortauthors{Wang et al.}

\begin{document}

\title{Photometry of Variable Stars from Dome A, Antarctica:\\
Results from the 2010 Observing Season}

\author{Lingzhi Wang\altaffilmark{1,2,3,4}, Lucas M.~Macri,\altaffilmark{2},
Lifan Wang\altaffilmark{2,3,5}, Michael C.~B.~Ashley\altaffilmark{6},\\Xiangqun
Cui\altaffilmark{3,7}, Long-Long Feng\altaffilmark{3,5}, Xuefei
Gong\altaffilmark{3,7}, Jon S.~Lawrence\altaffilmark{6,8}, Qiang
Liu\altaffilmark{3,4},\\Daniel Luong-Van\altaffilmark{6}, Carl
R.~Pennypacker\altaffilmark{9}, Zhaohui Shang\altaffilmark{3,10}, John
W.~V.~Storey\altaffilmark{6},\\Huigen Yang\altaffilmark{3,11}, Ji
Yang\altaffilmark{3,5}, Xiangyan Yuan\altaffilmark{3,7}, Donald
G.~York\altaffilmark{12}, Xu Zhou\altaffilmark{3,4},\\Zhenxi
Zhu\altaffilmark{3,5} \& Zonghong Zhu\altaffilmark{1,3}}

\altaffiltext{1}{Department of Astronomy, Beijing Normal University, Beijing, China}

\altaffiltext{2}{Mitchell Institute for Fundamental Physics \& Astronomy,
  Department of Physics \& Astronomy, Texas A\&M University, College Station,
  TX, USA}

\altaffiltext{3}{Chinese Center for Antarctic Astronomy, Nanjing, China}

\altaffiltext{4}{National Astronomical Observatory of China, Chinese Academy of
  Sciences, Beijing, China}

\altaffiltext{5}{Purple Mountain Observatory, Chinese Academy of
  Sciences, Nanjing, China}

\altaffiltext{6}{School of Physics, University of New South Wales, NSW, Australia}

\altaffiltext{7}{Nanjing Institute of Astronomical Optics and Technology,
  Nanjing, China}

\altaffiltext{8}{Australian Astronomical Observatory, NSW, Australia}

\altaffiltext{9}{Center for Astrophysics, Lawrence Berkeley National
  Laboratory, Berkeley, CA, USA}

\altaffiltext{10}{Tianjin Normal University, Tianjin, China}

\altaffiltext{11}{Polar Research Institute of China, Pudong, Shanghai,
  China}

\altaffiltext{12}{Department of Astronomy and Astrophysics and Enrico Fermi
  Institute, University of Chicago, Chicago, IL, USA}

\altaffiltext{}{Current email address: {\tt wanglingzhi@bao.ac.cn}}

\begin{abstract}
We present results from a season of observations with the Chinese Small
Telescope ARray (CSTAR), obtained over 183 days of the 2010 Antarctic winter.
We carried out high-cadence time-series aperture photometry of $\nbrs$~stars
with $i\lesssim15.3$~mag located in a 23 square-degree region centered on the
south celestial pole.

We identified $\nvar$~variable stars, including $\nnew$~new objects relative
to our 2008 observations, thanks to broader synoptic coverage, a deeper
magnitude limit and a larger field of view.

We used the photometric data set to derive site statistics from Dome A. Based
on two years of observations, we find that extinction due to clouds at this
site is less than 0.1 and 0.4~mag during 45\% and 75\% of the dark time,
respectively.
\end{abstract}


\section{Introduction}

Synoptic (time-series) astronomy has undergone a dramatic transformation over
the past decade thanks to advances in imaging and computer technology. Many
high-impact discoveries have taken place, such as transiting exoplanets
\citep{Charbonneau2000} and the detection of supernovae mere hours after
explosion \citep{Nugent2011}. These and other astrophysical problems benefit
from long, continuous and stable time-series photometry, which until recently
could only be achieved from space or via coordinated observations by a
world-wide telescope network. The former alternative is expensive, while the
latter is fraught with calibration issues, variable weather across the sites,
and is highly labor intensive.

One region on Earth -- the Antarctic Plateau -- offers an excellent alternative
to the aforementioned options by providing a combination of extended periods of
dark time, high altitude, low precipitable water vapor, extremely low
temperatures and a very stable atmosphere with greatly reduced scintillation
noise \citep{KenyonSL2006PASP-ScintillationDomeC}.  These conditions enable new
or extended observation windows in the infrared and sub-milimeter regions of the
electromagnetic spectrum, as well as improved conditions at optical and other
wavelengths. The resulting gains in sensitivity and photometric precision over
the best temperate sites can reach several orders of magnitude
(\citealt{Storey2005AS-AstronomyFromAntarctica,Storey2007CAA-Antarctic,
Storey2009AAPPS-AAA}; see also Tables 2 \& 4 and Figures 7-11 in
\citealt{Burton2010AAR-Antarctic}). The long ``winter night'' that occurs at these
latitudes and the minimal daily change in elevation for any given source make
the performance of a polar site equivalent to that of a six-site network at
temperate latitudes \citep{Mosser2007PASP-DutyCycle}. Some disadvantages include
the reduced fraction of the celestial sphere that can be observed, prolonged
twilight, and possibility of aurorae.

According to metereological studies carried out by
\citet{Saunders2009PASP-BestSite, Saunders2010EAS-BestSite}, the region
surrounding Dome A (elevation: 4,093 meters above mean sea level) in the
Antarctic plateau is likely the best astronomical site on Earth. In order to
further investigate the conditions at this promising site, we developed an
observatory capable of year-round operations called PLATO
\citep{Ashley2010HiA-PLATO, Luongvan2010SPIE-PLATO,
  Yang2009PASP-PLATO,Lawrence2009RSI-PLATO, Lawrence2008SPIE-PLATO,
  Hengst2008SPIE-PLATO,Lawrence2006SPIE-SiteTestA}, and a quad-telescope called
CSTAR \citep[the Chinese Small Telescope
  ARray,][]{Yuan2008SPIE-CSTAR,Zhou2010RAA-Testing}. The observatory is part of
the Chinese Kunlun station, located at Dome A.

Several papers were written based on a large amount of high-quality photometric
data obtained during the 2008 Antarctic winter: \citet{Zou2010AJ-Sky} undertook
a variety of sky brightness, transparency and photometric monitoring
observations; \citet{Zhou2010PASP-Catalog} published a photometric catalog of
$\sim10,000$~stars; and \citet{Wang2011} presented the first catalog of
variable stars in the CSTAR field of view.

\ \par

This paper presents an analysis of the data acquired by the CSTAR\#3 telescope
during the 2010 Antarctic winter season. \S2 briefly describes the
observations and data reduction; \S3 describes the steps followed to obtain
high-precision time-series photometry of the brightest $\nbrs$~stars with
well-sampled light curves; \S4 discusses the variable stars we discovered and
\S5 contains our conclusions.

\section{Observations and data reduction}

\subsection{Observations}

Observations were carried out using the same CSTAR telescope (unit \#3) which we
described in detail in \citet{Wang2011}. Briefly, it is a Schmidt-Cassegrain
wide-field telescope with a pupil entrance aperture of 145 mm, a Sloan $i$-band
filter, and a $1K\times1K$ frame-transfer CCD with a plate scale of
$15.7\arcsec/$~pix, giving a field of view $4.5^{\circ}$ on a side. The
telescope and camera have no moving parts, to allow for robust operation in the
extreme conditions of the Antarctic winter. Therefore, the field of view is
centered on the South Celestial Pole (SCP), which lies $\sim 9^{\circ}$ from the
zenith at this site, and exposures are short (20-40s) to prevent star
trails. The SCP field probes the inner halo of the Milky Way ($l=303^{\circ},
b=-27^{\circ}$) and has moderate extinction
\citep[$A_i=0.31$~mag,][]{Schlafly2011}.

Scientifically-useful images were acquired from 2010 March 29 to 2010
September 27. Table 1 lists the number of images and total integration time per
month and Table 2 lists the different integration times used throughout the
observing season. More than 338,000 images (equivalent to over 360 GB of data)
were collected with a total integration time of 2,553 hours.

\subsection{Data Reduction}
\label{sec:proc}
The preliminary data reduction for raw science frames involved bias
subtraction, flat fielding, correction for variations in sky background,
fringe pattern subtraction and bad pixel masking. We used the same bias frame
from our previous paper, which was created during instrument testing in
China. We generated a sky flat by median-combining 2,700 frames with high sky
level ($>$ 10,000 ADU) taken throughout the observing season during twilight
conditions (Sun elevation angle between 0 and $-10^{\circ}$). Prior to
combining the images, we masked any stars present in the individual frames
using a detection threshold of $2\sigma$, as well as regions within 10 pixels
of any pixel close to saturation (defined as $>$25,000 ADU).

Approximately 35\% of the bias-subtracted and flat-fielded images showed
low-frequency sky background variations with an amplitude of $\sim 3$\% of the
mean value, which we subtracted by the following method. We divided each image
into 1,024 square sections (32 pixels on a side) and computed the mean and
standard deviation of the sky value in each section using the procedure
implemented in DAOPHOT \citep{Stetson1987}. We did not use any sections with
sky values above 20,000 or below 0 ADU, or those with standard deviations
that exceeded 300 ADU. Those properties are typical of very crowded regions
of an image, or indicate the presence of a very bright star, so we replaced
the sky value of those sections with the median value of their nearest ten
neighbors. We

\begin{deluxetable}{lrr}
\tablewidth{0pt}
\tablecaption{Log of observations\label{tb10:log}}
\tablehead{\colhead{Month} & \colhead{\# images} &
\colhead{Total exp.}\\ \colhead{2010} & & \colhead{time (hr)}}
\startdata
March       &   1587 &   17.6 \\
April       &  31110 &  345.7 \\
May         &  39651 &  405.8 \\
June        &  69509 &  579.2 \\
July        &  97310 &  631.1 \\
August      &  73088 &  406.0 \\
September   &  30098 &  167.2 \\
{\bf Total} &{\bf 342353} & {\bf 2552.6}
\enddata
\end{deluxetable}

\begin{deluxetable}{llcr}
\tablewidth{0pt}
\tablecaption{Exposure times\label{tb10:exp}}
\tablehead{\colhead{Start} & \colhead{End} & \colhead{Exp.} & \colhead{\#}\\
\colhead{date} & \colhead{date} & \colhead{(s)} & \colhead{images}}
\startdata
2010-03-29 & 2010-05-26 & 40 &  59843 \\
2010-05-26 & 2010-07-13 & 30 & 114577 \\
2010-07-13 & 2010-09-27 & 20 & 167933
\enddata
\end{deluxetable}

\noindent{\\ generated a full-resolution background model frame by
  fitting the 1,024 sky values using a thin-plate spline. We
  calculated the standard deviation of sky values before and after
  model subtraction; if the reduction in the scatter was less than a
  factor of 2, we reduced the section size to 16 pixels on a side and
  repeated the procedure.}
 
The background-subtracted images exhibited a fringe pattern, which is commonly
encountered in CCD images obtained at near-infrared wavelengths \citep[for a
review, see][]{howell2012}.  We created a fringe correction frame and removed
this instrumental signature from our images following these steps. First, we
selected 1,500 images taken each month with the lowest sky levels during dark
time (Sun elevation angle below $-18^{\circ}$) and 1,500 images taken during
twilight. This yielded 7 sets of images obtained during twilight and 5 sets
acquired during dark time (April to August). The motivation behind creating the
two different sets was to check for variations in fringe amplitude as a
function of Sun elevation angle.

Next, we removed stars by masking any pixel lying more than $2\sigma$ above the
sky level. We combined the frames in each of the 12 sets by sorting the
un-masked pixels at each $(x,y)$ position, discarding the lowest 10 values
(where the fringe pattern was too low to be useful), and averaging the next 50
values. The fringe pattern had an average peak-to-peak amplitude of 2\% of the
mean sky value, with no statistically significant variation as a function of
time or Sun elevation angle. Therefore, we generated the final fringe frame by
taking the minimum pixel value at each location among the 12 sets (to further
ensure that no stellar sources were included). The fringe pattern was
subtracted from individual frames by iteratively scaling the correction image
until the sky background exhibited the lowest {\it r.m.s.}\ value.

\begin{figure}[t]
\begin{center}
\includegraphics[width=0.49\textwidth]{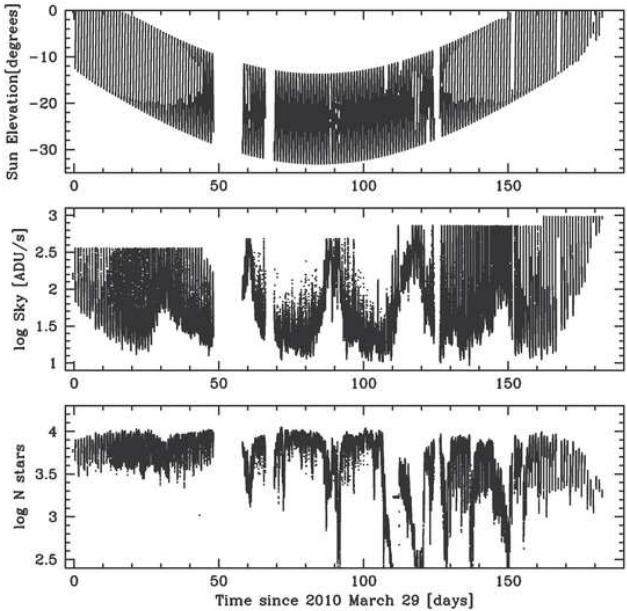}
\caption{\label{fig:info} Time-series plots of Sun elevation angle (top panel), sky brightness level (middle panel), and number of stars detected in each image (bottom panel).\vspace*{-12pt}}
\end{center}
\end{figure}  

\section{Photometry}

\subsection{Frame Selection and  Photometry}
We carried out initial photometric measurements on the debiased, flattened,
sky-subtracted and fringe-corrected images using the same pipeline and
procedures described in \citet{Wang2011}, which we briefly summarize
here. Given the extremely undersampled nature of the images, we performed
aperture photometry using DAOPHOT \citet{Stetson1987}. We set the aperture
radius to 2.5 pixels (equivalent to $39.3\arcsec$) and the sky annulus extended
from 4 to 7 pixels (equivalent to $62.8-109.9\arcsec$). The detection threshold
was set to $5\sigma$ above the sky background, which typically corresponded to
$i\sim14$~mag.

Figure~\ref{fig:info} displays the Sun elevation angle (top panel), sky
background in ADU s$^{-1}$ (middle panel), and number of stars detected (bottom
panel) for each image as a function of time since 2010 March 29. The sky
background shows a clear monthly pattern related to the lunar phase. During
most of the winter season, the data acquisition system was programmed to stop
when the sky level exceeded 14,600 ADU; this limit was raised to 19,400 after
2010 Sep 7. We initially selected frames with Sun elevation angle below
$0^{\circ}$ and more than 1,500 stars. These criteria were met by 85.5\%
(289,343) of all the images acquired during the Antarctic winter. A period of
bad weather and ice buildup on the top cover of the instrument during the month
of July resulted in the rejection of $\sim$35,000 images (equivalent to 8 days
of operations).

The selected images have a median sky level of 32 ADU s$^{-1}$, equivalent to a
median sky background of $i\!=\!19.8$~mag$/\sq \arcsec$. This value is identical
to the value derived by \citet{Zou2010AJ-Sky} and similar to our previous
determination of 19.6 mag$/\sq \arcsec$ \citep{Wang2011}. Note that all these
estimates include contributions from moonlight. The darkest sky background
measured during the season (on clear moonless nights) was $i\sim20.9$~mag$/\sq
\arcsec$. The median value of the number of stars detected in an individual
frame is 8,600, higher than the corresponding value of 7,500 from the 2008
observations. We attribute this increase to the subtraction of sky background
variations described in \S~\ref{sec:proc}, which enabled the detection of
fainter stars at a fixed threshold.

\begin{figure}
\begin{center}
\includegraphics[width=0.49\textwidth]{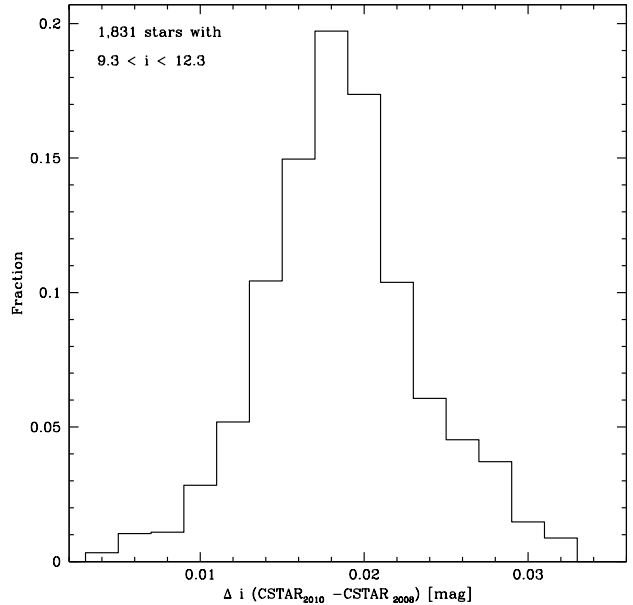}
\caption{Comparison of the mean instrumental magnitudes of bright stars in common between
  the 2008 and 2010 CSTAR observations.\label{fig:zpt}}
\end{center}
\end{figure}

We used the initial photometry to register all frames using DAOMATCH and
DAOMASTER and created a reference image (hereafter, ``master frame'') with
$4\times$ finer spatial sampling. We followed the same procedure described in
our previous paper, this time based on 2,580 high-quality images obtained
during 2010 June 13. We carried out aperture photometry on the master frame
using the same parameters listed above and detected approximately 155,000
stars, reaching a depth of $i\sim20.9$~mag. Additionally, we performed
point-spread function photometry that, despite its lower quality due to the
undersampled nature of the images, enabled us to estimate the relative
magnitudes of stars with overlapping apertures.

\subsection{Photometric corrections}

We performed all the photometric corrections described in detail in
\citet{Wang2011}, which included: exposure time normalization, zeropoint
correction, sigma rescaling, residual flat fielding, time calibration, masking
of satellite trails and saturated regions, spike filtering, and magnitude
calibration. We will only discuss the details of magnitude and time
calibration since the rest of the procedures were identical to our previous
work.

We performed the magnitude calibration by matching the mean instrumental
magnitudes of 1,831 stars with $9.3 \leq i \leq 12.3$~mag in common between our
2010 and 2008 master frames. Since the mean instrumental magnitudes of each
season are referenced to individual images obtained under the best observing
conditions (in terms of sky background and number of stars) we would expect
them to reflect photometric conditions and therefore have very similar
zeropoints. Indeed, we found a zeropoint offset of only $0.018\pm 0.005$~mag as
shown in Fig.~\ref{fig:zpt}. We transformed the instrumental magnitudes to
the Sloan photometric system by applying the offset of $7.45\pm0.04$~mag
previously derived in \S3.2 of \citet{Wang2011}. That offset was based on the
comparison of our photometry with the $i$ synthetic magnitudes derived by
\citet{Ofek2008} using the {\it Tycho} catalog, which carries an additional
systematic uncertainty of 0.02~mag.

\begin{figure}[t]
\begin{center}
\includegraphics[width=0.49\textwidth]{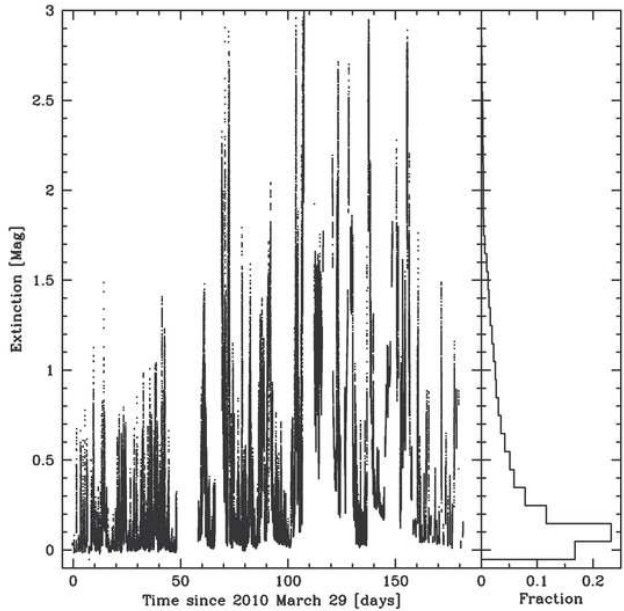}
\caption{Left panel: time series plots of differential extinction, right
panel: distribution of differential extinction. \label{fig:difext}\vspace*{-12pt}}
\end{center}
\end{figure}

Figure~\ref{fig:difext} shows a time series and a histogram of differential
extinction values, based on the photometry procedures described above. We find
that extinction due to clouds in the $i$ band at Dome A is less than 0.4 mag
during 70\% of the dark time, and less that 0.1 mag during 40\% of the dark
time.  These values are similar to those previously derived for the 2008
Antarctic winter season \citep[80\% and 50\%, respectively, from][]{Wang2011}.

The computer associated with the CSTAR\#3 telescope has a GPS receiver to
maintain time synchronization, and this time was to be distributed to
computers controlling the other CSTAR telescopes (including the one used for
these observations). However, there was a communication problem between the
computers throughout the entire observation period, which led to a drift in
the time stamp of the FITS images. \S4.3 of \citet{Zhou2010PASP-Catalog}
contains details of the time calibration for the 2008 data; we carried out a
similar procedure for the 2010 data as explained below.

\begin{figure}[t]
\begin{center}
\includegraphics[width=0.49\textwidth]{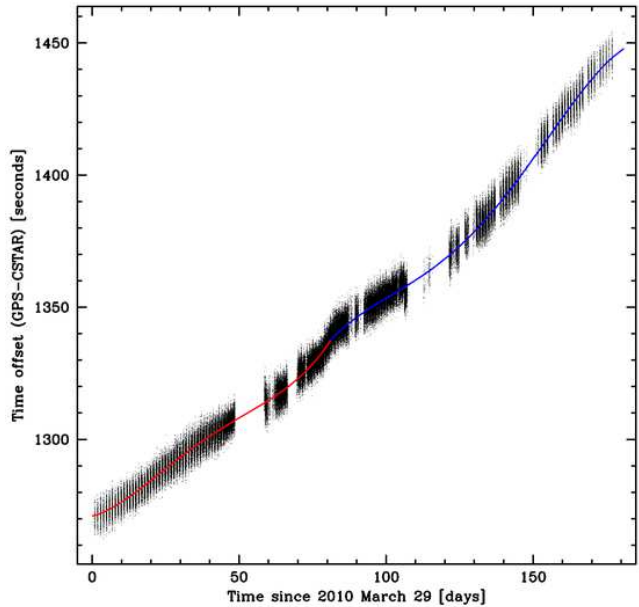}
\caption{Drift between true local time and computer time, fitted by two 
fourth-order polynomials.\label{fig:timecal}}
\end{center}
\end{figure}

We performed the time calibration in three steps. First, we identified two
bright stars located very close to R.A.=0h and measured the angle (with respect
to the x-axis) of a line extending from the SCP to their positions. We
calculated the difference between the measured angle and the one predicted from
the time in the FITS headers, and fit a fourth-order polynomial to solve for the
time drift. The smallest dispersion was obtained by fitting two different
polynomials for the data acquired before and after 2010 June 17. The total clock
drift over 6 months amounted to $\sim 190$~seconds as seen in
Fig.~\ref{fig:timecal}. Next, we used 19 entries in a log (obtained for
engineering purposes throughout the season) that listed the time offset between
the CSTAR\#3 computer and another computer at Dome A which had maintained GPS
synchronization. While these data points were not sufficient to solve for the
time drift, they served to transform our relative measurements into an absolute
reference frame. We obtained an offset of $1337.5\pm1.6$~s. Lastly, we applied a
transformation to the heliocentric reference frame.

We later checked the derived time offset by performing a cross-correlation of
the light curves of 33 bright, high-amplitude, periodic variables (such as
eclipsing binaries and RR Lyraes) that we had previously detected in
\citet{Wang2011}. We used the latest implementation of the Phase Dispersion
Minimization algorithm \citep{Stellingwerf1978, Stellingwerf2011}, which can
robustly handle large gaps in the time series and combines all light curve
``segments'' to determine a common period. Using this approach, we found a
consistent (but less precise) offset of $1270\pm245$~s.

\section{Variable Stars in the CSTAR field}

While the primary goal of CSTAR observations is to characterize the observing
conditions at Dome A, the rapid cadence of image acquisition and the long
duration of the winter night make this a relatively unique data set for
studying variable stars and searching for extrasolar planets via the transit
method. We used several complementary techniques (described below) to identify
variables among the selected stars, since no single approach is sensitive to
all types of stellar variability. Some of the methods we used are only
sensitive to periodic variability. Most of the algorithms we used are
available as part of the {\tt VARTOOLS} light curve analysis program of
\citet{hartman08}.

Having already characterized the observing conditions at the site based on all
the ``dark time'' data, we restricted our variable-star analysis to the images
obtained under the best observing conditions, which we defined as a sky
background below 100 ADU s$^{-1}$ and an extinction $\leq0.5$~mag. These
criteria were met by $169{,}500$ frames (the ``science-quality'' sample),
corresponding to about 60.9\% of the previously selected set of images. Among
the rejected images, 3.7\% failed to provide a reliable coordinate
transformation with respect to the reference frame.

\begin{figure}[t]
\begin{center}
\includegraphics[height=0.49\textwidth]{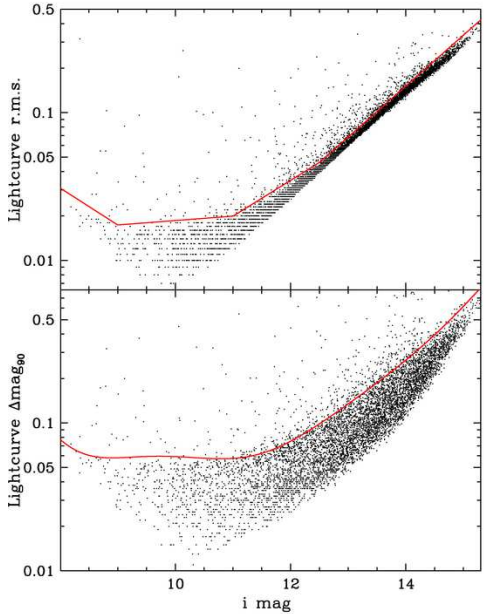}
\caption{{\it r.m.s.}\ (top) and 90\%-ile magnitude range (bottom) of the
  lightcurves of $\nbrs$~selected stars with sufficient synoptic coverage (see
  text for details).\label{fig:magerr}}
\end{center}
\end{figure}
 
We initially selected the brightest $\nbrt$~stars in the master frame
(corresponding to a depth of $i\sim15.3$~mag) for time-series aperture
photometry. Once the measurements were carried out, we restricted our analysis
to $\nbrs$~objects with valid measurements in at least 20\% of the individual
science-quality frames (or in 20\% of all possible 3,000s time segments) in
order to ensure sufficient synoptic coverage. This restriction implies a
maximum declination limit of $-87^\circ 13\arcmin$ for objects in the sample,
which we will later use when comparing our variable star statistics with
previous studies.

We complemented the search for variables with information obtained from the
PSF photometry previously carried out on the master frame, which had a
$4\times$ finer pixel scale and represented a 24-hour average stellar
flux. For each object of interest, we calculated the fraction of contaminating
flux contributed all other stars located closer than a certain distance: the
aperture radius ($39.3\arcsec$), the inner edge of the sky annulus
($62.8\arcsec$), and the outer edge of the sky annulus ($109.9\arcsec$). We
refer to these three fluxes hereafter as ``close'', ``medium'' and ``far''
respectively.

\begin{figure}[t]
\begin{center}
\includegraphics[width=0.49\textwidth]{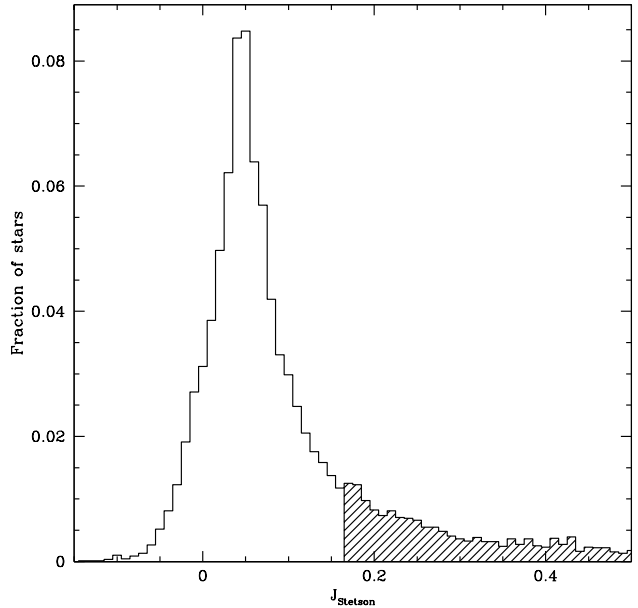}
\caption{\label{fig:jste} Variability statistic $J$ \citep{Stetson1996} for
  $\nbrs$~selected stars with sufficient synoptic coverage (see text
  for details).}
\end{center}
\end{figure}
 
\subsection{Search for variability}
 
The first phase of our search for variable stars was aimed at identifying
objects with statistically significant variations in magnitude that did not
necessarily display a periodic behavior during our observations -- including
objects such as Miras, other very long-period variables and irregular
variables. We used a combination of three metrics to separate constant from
variable stars, as detailed below.

We calculated the {\it r.m.s.}\ and the magnitude range spanned by 90\% of data
points (hereafter, $\Delta i_{90}$) of the light curves of all stars. We
determined upper $2\sigma$ envelopes for both quantities as a function of
magnitude, shown with red lines in Fig.~\ref{fig:magerr}. Stars lying above
both envelopes were flagged as possible variables for further analysis. The
upturn in both envelopes for $i<9$~mag is probably due to a combination of
factors, including the onset of non-linearity in the detector and the dearth of
truly constant stars in this magnitude range within our field. The Besan\c{c}on
model of the Galaxy \citep{Robin2003} predicts that post-main sequence stars
\citep[which are very likely to exhibit variability, see][]{Henry2000} will
outnumber main-sequence objects by a ratio of 7 to 1 at this magnitude.

The lowest light curve {\it r.m.s.}\ values (7 mmag) are found at $i\sim10$~mag,
reaching within a factor of 5 of the expected scatter due to scintillation
\citep{Young1967} for our telescope at Dome A. Given the large number of
individual measurements in our photometry, bright ($i<13.5$~mag) stars lying
below both envelopes have statistical uncertainties in their mean magnitudes
below $2\times 10^{-4}$~mag. As discussed in \S3.2 and in \citet{Wang2011}, the
overall uncertainty of the photometry is dominated by the statistical ($\pm
0.04$~mag) and systematic ($\pm 0.02$~mag) uncertainties of the calibration
procedure.

Next, we computed the Welch-Stetson variability index $J$ \citep[\S2
of][]{Stetson1996}, including the usual rescaling of the reported DAOPHOT
magnitude errors \citep{udalski94,kaluzny98}. The result of this analysis is
shown in Fig.~\ref{fig:jste}; as expected for the $J$ statistic, there is a
gaussian distribution of values with a mean value close to zero (corresponding
to stars with no significant variability), and a one-sided tail (towards
positive values) of candidate variable stars. We determined a mean value of
$J=0.049\pm0.038$ by fitting a Gaussian function to all objects with
$|J|<0.3$, and flagged stars with $J \geq 0.164$ (equivalent to a
$+3\sigma$ selection) for further inspection.

We combined the {\it r.m.s.}, $\Delta i_{90}$ and $J$ criteria listed above
to select 44 variable stars. In addition to passing all 3 variability
criteria, the selected objects were also restricted to have contaminating
fluxes from nearby companions in the ``medium'' and ``far'' apertures below
11.5\% and 23\% of the total flux, respectively. These limits may have
resulted in the rejection of some {\it bona fide} aperiodic variables or
transient events, but they serve to eliminate false positives from our
sample.

\subsection{Search for periodic variability}

The $J$ statistic was designed to be sensitive to statistically significant
photometric variability between neighboring data points, and is well suited to
detect continuously-varying objects such as pulsating stars (Miras, Cepheids, RR
Lyrae, $\delta$~Scuti, etc.) or contact binaries. It is not particularly
sensitive to detached eclipsing binaries (where the variation is restricted to
a very small fraction of the phase) or to objects where the variation only
becomes statistically significant after phasing many cycles (such as
transiting exoplanets or very low amplitude pulsators). Similarly, the other
two metrics used in the first phase of the search for variables (light curve
{\it r.m.s.}\ and $\Delta i_{90}$) lack sensitivity to small-scale periodic
variations.  Therefore, we searched for {\it periodic} variability among the
stars that had failed one or more of the previous selection criteria using two
of the techniques described in detail in \citet{Wang2011}: the Lomb-Scargle
method \citep[][hereafter LS]{Lomb1976, Scargle1982} and the ``box fitting
algorithm'' of \citet[][hereafter BLS]{Kovacs2002}. 

Given the design of the CSTAR system, stars describe daily circular tracks
through the CCD. This can lead to spurious detections of periodicity due to
small residual flat-field variations. Fortunately, these false positives are
easily identified and discarded because the variations occur at frequencies
close (0.5-3\%) to 1 cycle per sidereal day and their (sub-)harmonics. We
considered an object to have significant periodic variability if the period
determined by VARTOOLS (based on either the LS or the BLS technique) had a
SNR$>30$ and lied outside of the excluded frequencies. We imposed a minimal
restriction on contaminating flux for this search, removing 2\% of objects
where the contaminating flux from other stars in the ``close'' aperture
exceeded 40\% of the total. The search for periodic variables can tolerate
substantially larger contamination from neighbors for several reasons: very
close neighbors (within the instrumental PSF or the measurement aperture) can
only decrease the amplitude of the variation but cannot result in a false
positive, and future observations can be used to identify which of the
confused stars is the actual variable; stars outside the measurement aperture
but within the outer sky annulus may produce noisier measurements but the long
span of our observations still yields high-quality binned light curves that
satisfy the requirement on the SNR of the period determination; any
companions that somehow induce a spurious periodic variation will do so at
frequencies of 1 cycle/day or one of its harmonics and those have already been
removed from consideration.

We detected an additional 136 variables using the LS technique and a further 8
variables using the BLS technique. These algorithms were also applied to the
variables previously selected in \S4.1 and 16 of those objects were found to
exhibit a significant periodicity. The initial periods were refined using the
Period04 program \citep{Lenz2005} and the Phase Dispersion Minimization
algorithm \citep{Stellingwerf1978, Stellingwerf2011}.

Approximately two thirds of the periodic variables exhibit a highly regular
variation both in terms of period and amplitude (e.g., eclipsing binaries,
normal RR Lyraes), a few have a long-term modulation in amplitude (e.g.,
Blazhko RR Lyraes), and about one third show variations at multiple
frequencies. In the latter two cases, we based our analysis in the most
significant period.

\subsection{Properties of variable stars}

\begin{figure*}[h]
\begin{center}
\includegraphics[height=\textwidth]{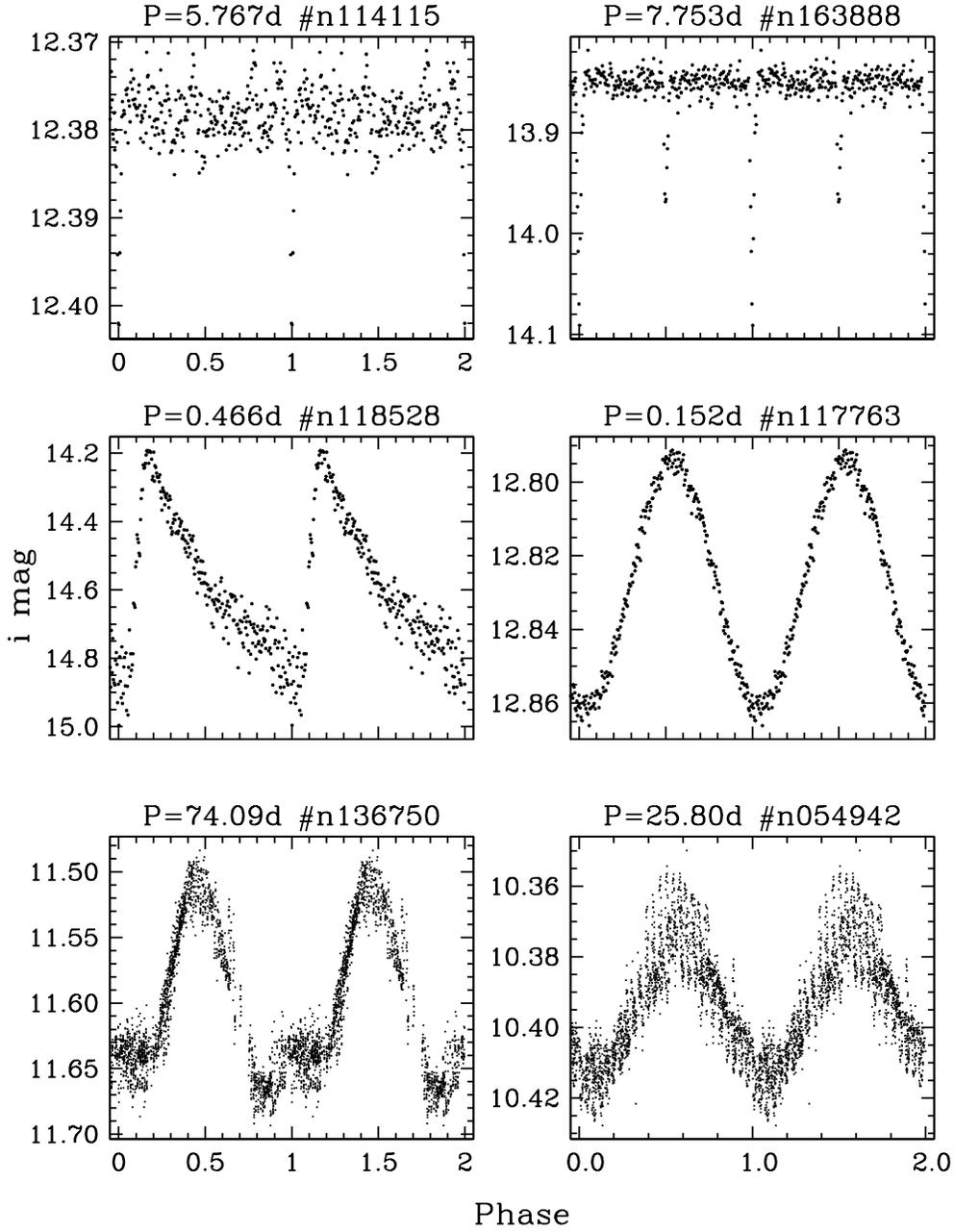}
\caption{Phased light curves of six newly detected periodic variable
  stars. The periods and 2010 CSTAR IDs are given above each panel. Top row:
  transiting exoplanet candidate and detached eclipsing binary; middle row: RR
  Lyrae and $\delta$~Scuti; bottom row: multi-periodic variables phased at the
  most significant period.\label{fig:lcphas}}
\end{center}
\end{figure*}

\begin{figure*}[h]
\begin{center}
\includegraphics[width=\textwidth]{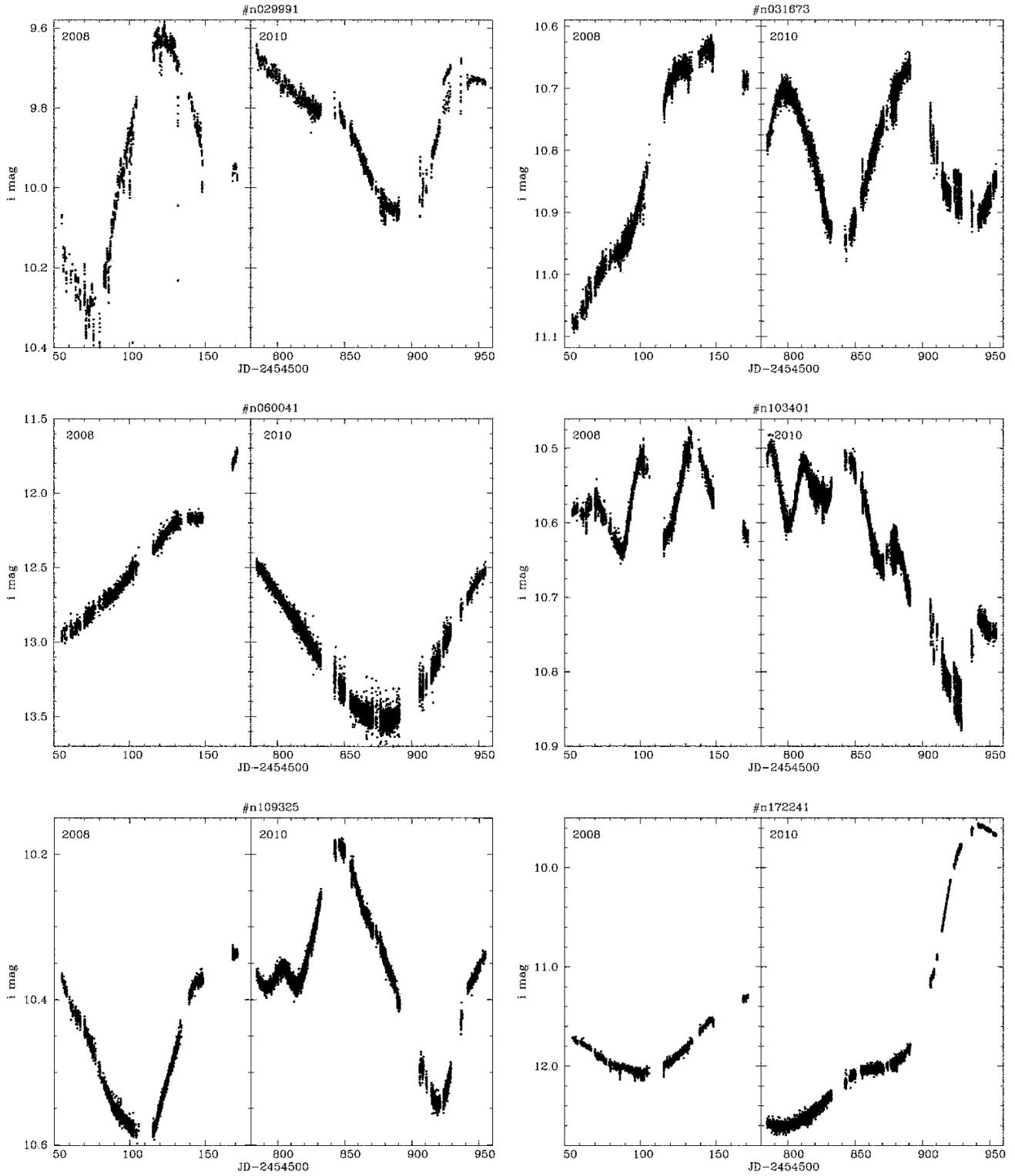}
\caption{Two-year light curves (left: 2008; right: 2010) for variables with
  long-term trends.}
\label{fig:lctime}
\end{center}
\end{figure*}

Table~\ref{tab:variables} lists the properties of all detected variables.
Column 1 has the 2010 CSTAR ID (a letter {\it n} is added at the beginning to
avoid confusion with 2008 CSTAR IDs from our previous work); column 2 lists the
2008 CSTAR ID from \citep{Wang2011}, if applicable; column 3 presents the ID
from the Guide Star Catalog, version 2.3.2 (GSC2.3), when available; columns 4
and 5 give the right ascension and declination; column 6 contains the mean
$i$-band magnitude; column 7 lists the 90\% range of the $i$-band light curve;
column 8 has the $J$ value; column 9 lists the most significant period (when
applicable); column 10 specifies the technique used to find the period; column
11 gives the first time of minimum light contained in our observations (listed
only for the periodic variables); column 12 contains a tentative classification
of the variable type, when possible; column 13 has additional information,
including previous identification of the variables by the All-Sky Automated
Survey \citep[ASAS,]{Pojmanski2005} or inclusion in the General Catalogue of
Variable Stars \citep[GCVS,][]{Samus2009}. Representative light curves of
periodic variables are shown in Fig.~\ref{fig:lcphas}, while two-season
light curves of selected long-term variables are plotted in
Fig.~\ref{fig:lctime}. Detailed finding charts including light curve plots
for all variables are available through the Chinese Virtual
Observatory\footnote{ \url {http://casdc.china-vo.org/data/cstar}}. All light
curve data is also available through this site.

We detected a total of $\nvar$ variables in the 2010 CSTAR data, consisting of
$\nnew$ new objects and $\nold$ stars in common with \citet{Wang2011}. We did
not recover 36 objects classified as variables in our previous paper for the
following reasons: 10 did not have enough data points, 7 were blended, and 19
did not meet the selection criteria adopted in this paper. The new variables
detected in this study relative to our previous work were made possible by the
deeper magnitude limit and slightly larger field of view ($\delta
<-87^{\circ}13\arcmin$) of the 2010 observations, which were respectively due
to better sky subtraction and improved alignment of the center of the field
with the SCP.

Thanks to the greater depth and synoptic\ \  coverage\ \  of\ \  our\ \  observations,\ \  we obtained\ \  a\ \  $\sim4\times$\ \  increase\ \  in \ \ the number\ \  of\ \  variables\ \  with\ \  $\delta < -87^{\circ}13\arcmin$\ \ relative\ \ to 

\clearpage

\LongTables
\begin{deluxetable*}{lllllrrrrlrll}
\tabletypesize{\tiny}
\tablewidth{0pt}
\tablecaption{Variable stars\label{tab:variables}}
\tablehead{
\multicolumn{3}{c}{ID}                                                       & \multicolumn{1}{c}{R.A.} & \multicolumn{1}{c}{Dec.}  & \multicolumn{1}{c}{$i$}& \multicolumn{1}{c}{$\Delta i_{90}$} &  \multicolumn{1}{c}{$J$} & \multicolumn{2}{c}{Period}                              & \multicolumn{1}{c}{$T_0^3$} & \multicolumn{1}{c}{Class$^{4}$} & \multicolumn{1}{c}{Note$^{5}$}\\
\multicolumn{1}{c}{2010} & \multicolumn{1}{c}{2008} &\multicolumn{1}{c}{GSC} & \multicolumn{2}{c}{(J2000.0)$^{1}$}                  &\multicolumn{2}{c}{(mag)}                                     &                          & \multicolumn{1}{c}{(d)} & \multicolumn{1}{c}{Src$^{2}$} & \multicolumn{1}{c}{(d)}     &                                 &                              }
\startdata
n010320 &    \nd & S742000016 & 12:04:48.75 & $-$87:23:06.0 &  8.79 &  0.18 &  9.68 &    \nd    & \nd &   \nd    & IR & A   \\ 
n012443 & 001707 & S74D000321 & 12:32:42.91 & $-$87:26:22.9 & 11.10 &  0.55 &  9.95 &  0.338544 &  LS & 785.6824 & EC & A   \\ 
n012506 &    \nd & S3Y9000067 & 11:44:32.67 & $-$87:27:35.9 & 10.89 &  0.12 &  2.07 & 31.447988 &  LS &   \nd    & MP &     \\ 
n015318 & 003125 & S3YM000469 & 10:43:46.63 & $-$87:25:10.1 &  9.69 &  0.09 &  3.32 &  3.602742 &  LS &   \nd    & MP &     \\ 
n015705 &    \nd & S3YM000358 & 10:04:32.92 & $-$87:13:44.7 & 10.97 &  0.08 &  0.65 & 19.081678 &  LS &   \nd    & MP &     \\ 
n016257 & 003697 & S742000061 & 12:08:11.93 & $-$87:35:39.9 & 11.56 &  0.08 &  0.75 & 26.567900 &  LS &   \nd    & MP &     \\ 
n016505 & 003850 & S742000043 & 12:34:25.12 & $-$87:34:37.7 & 10.14 &  0.06 &  1.59 & 17.251594 &  LS &   \nd    & MP & A   \\ 
n017573 & 004463 & S3YM000518 & 10:40:16.05 & $-$87:29:29.8 & 11.16 &  0.06 &  0.58 &  0.869262 &  LS & 785.7941 & ED &     \\ 
n017781 &    \nd & S74D000351 & 13:21:24.66 & $-$87:29:48.4 & 12.41 &  0.10 &  0.25 &  4.822329 &  LS &   \nd    & MP &     \\ 
n020508 &    \nd & S74D000440 & 13:01:58.40 & $-$87:39:56.3 &  9.01 &  0.16 &  3.11 &  5.798380 &  LS & 786.5008 & ED &     \\ 
n023757 &    \nd & S3YM000018 & 09:51:32.15 & $-$87:28:32.7 & 13.24 &  0.18 &  0.20 &  0.240900 &  LS & 785.4140 & PR &     \\ 
n024696 & 009171 & S3YM000662 & 10:12:54.85 & $-$87:38:22.9 & 14.10 &  0.77 &  1.29 &  0.591726 &  LS & 785.4091 & RL & A   \\ 
n025073 &    \nd & S3YN000420 & 09:29:27.99 & $-$87:21:39.6 &  9.01 &  0.32 & 18.03 &    \nd    & \nd &   \nd    & IR & A   \\ 
n025734 & 009952 & S742000182 & 12:43:30.67 & $-$87:53:30.9 & 11.30 &  0.20 &  2.89 & 23.819339 &  LS &   \nd    & MP & A   \\ 
n027942 & 011616 & S3Y9000240 & 10:56:28.86 & $-$87:55:21.0 & 11.97 &  0.08 &  0.55 & 27.282715 &  LS &   \nd    & MP &     \\ 
n028073 & 011709 & S742000246 & 12:41:44.27 & $-$87:58:28.5 & 11.37 &  0.07 &  0.77 &  2.951167 &  LS & 787.1734 & PR &     \\ 
n028235 & 011796 & S742000286 & 12:21:35.82 & $-$88:00:14.5 & 12.20 &  0.26 &  0.90 &  1.892910 &  LS & 786.9435 & ES &     \\ 
n029044 &    \nd & S742000186 & 13:28:28.82 & $-$87:53:09.2 &  9.52 &  0.04 &  1.86 & 11.564653 &  LS &   \nd    & MP &     \\ 
n029991 & 013255 & S74F000377 & 14:54:21.37 & $-$87:21:05.0 &  9.87 &  0.34 &  6.41 &    \nd    & \nd &   \nd    & IR & A   \\ 
n030008 & 013140 & S3Y9000236 & 10:25:53.99 & $-$87:53:40.8 &  9.83 &  0.07 &  1.40 & 20.978237 &  LS &   \nd    & MP &     \\ 
n031343 & 014111 & S74F000634 & 14:29:01.63 & $-$87:38:16.2 & 13.74 &  0.42 &  0.67 &  0.174082 &  LS & 785.3263 & DS &     \\ 
n031673 & 014368 & S3YM000753 & 10:03:10.79 & $-$87:51:06.2 & 10.81 &  0.25 &  5.38 & 95.147196 &  LS &   \nd    & MP &     \\ 
n031801 & 014495 & S3Y9000339 & 10:49:00.65 & $-$88:02:17.2 & 12.30 &  0.12 &  0.65 & 19.218473 &  LS &   \nd    & MP &     \\ 
n034888 & 016836 & S742030458 & 12:49:16.22 & $-$88:11:17.6 & 13.74 &  0.30 &  0.47 &  0.176214 &  LS & 785.2034 & DS &     \\ 
n035474 &    \nd & S74F007170 & 15:09:55.59 & $-$87:25:01.9 &  9.09 &  0.08 &  2.89 &    \nd    & \nd &   \nd    & IR &     \\ 
n037305 & 018708 & S3YN000609 & 09:02:20.82 & $-$87:37:41.0 & 12.38 &  0.09 &  0.25 &  1.627880 &  LS & 785.2541 & ES &     \\ 
n039537 & 020436 & S3YN000517 & 08:40:33.28 & $-$87:28:38.6 & 12.60 &  0.23 &  0.76 & 61.305131 &  LS &   \nd    & MP &     \\ 
n039664 & 020526 & S742000504 & 13:23:49.26 & $-$88:16:04.3 & 12.46 &  0.42 &  1.53 &  2.510726 &  LS & 786.9545 & ES & A   \\ 
n040035 &    \nd & S3Y9000616 & 11:52:51.12 & $-$88:23:28.9 &  9.21 &  0.06 &  1.73 & 17.086312 &  LS &   \nd    & MP &     \\ 
n042221 & 022489 & S3Y9000527 & 10:01:21.80 & $-$88:13:30.8 & 11.89 &  0.22 &  2.51 &  0.652314 &  LS & 785.7043 & EC & A   \\ 
n043406 &    \nd & S3YN000632 & 08:39:40.85 & $-$87:39:02.3 & 12.21 &  0.10 &  0.26 &  7.165431 & BLS & 787.9603 & ED &     \\ 
n043566 & 023614 & S742000638 & 12:09:34.38 & $-$88:29:59.0 & 11.49 &  0.05 &  0.32 & 17.256627 &  LS &   \nd    & MP &     \\ 
n044666 &    \nd & S3Y9000612 & 10:32:32.73 & $-$88:25:02.6 & 13.73 &  0.14 &  0.09 &  3.535402 & BLS & 788.5108 & ED &     \\ 
n045656 &    \nd & S742000672 & 14:20:52.04 & $-$88:14:33.4 & 12.65 &  0.08 &  0.25 &  0.400924 &  LS & 785.2228 & EC &     \\ 
n045983 & 025440 & S742000656 & 12:12:56.28 & $-$88:34:11.6 & 10.77 &  0.03 &  0.37 &  9.972472 &  LS &   \nd    & MP &     \\ 
n046511 & 025846 & S742000671 & 14:25:39.12 & $-$88:14:54.1 & 13.77 &  0.14 &  0.07 &  9.618211 &  LS & 788.7800 & PR &     \\ 
n047552 & 026640 & S3YB000429 & 11:17:00.40 & $-$88:35:36.4 &  9.37 &  0.38 & 10.24 &    \nd    & \nd &   \nd    & LT & A   \\ 
n047660 & 026730 & S3YB000202 & 09:58:36.02 & $-$88:23:59.9 & 12.89 &  0.11 &  0.13 &  2.066699 &  LS & 786.2584 & ED &     \\ 
n049496 & 028221 & S3YB000436 & 11:04:06.48 & $-$88:38:02.5 & 13.55 &  0.25 &  0.38 &  4.815779 &  LS & 788.6982 & PR &     \\ 
n050944 & 029379 & S74E000029 & 15:57:05.68 & $-$87:30:05.4 & 11.93 &  0.15 &  0.99 &  1.553117 &  LS &   \nd    & MP & A   \\ 
n052190 & 030353 & S3YA000615 & 08:39:40.45 & $-$88:03:19.4 & 12.87 &  0.09 &  0.15 &  1.048425 &  LS & 785.9799 & PR &     \\ 
n052325 &    \nd & S74E000072 & 15:44:44.17 & $-$87:46:35.4 &  8.35 &  0.09 &  5.76 &    \nd    & \nd &   \nd    & IR &     \\ 
n054284 & 032007 & S3YB000225 & 09:29:39.14 & $-$88:30:06.4 & 11.78 &  0.12 &  0.98 &  0.621863 &  LS & 785.5950 & RL & A,G \\ 
n054942 & 032544 & S3Y8000312 & 11:48:09.42 & $-$88:49:52.6 & 10.39 &  0.05 &  1.27 & 25.795462 &  LS &   \nd    & MP &     \\ 
n055150 &    \nd & S3YB000458 & 10:04:40.34 & $-$88:40:25.5 &  9.06 &  0.04 &  1.41 &  0.092508 &  LS & 785.2723 & PR &     \\ 
n057617 & 034669 & S743000094 & 13:50:03.38 & $-$88:46:13.2 & 10.08 &  0.06 &  1.40 & 15.167192 &  LS &   \nd    & MP &     \\ 
n057725 & 034724 & S3YB000482 & 10:01:18.94 & $-$88:44:36.8 & 10.08 &  0.10 &  2.34 & 43.205800 &  LS &   \nd    & MP & A   \\ 
n057789 &    \nd & S3Y8000318 & 11:19:13.27 & $-$88:53:40.8 & 12.74 &  0.10 &  0.19 &  1.862438 &  LS &   \nd    & MP &     \\ 
n058002 & 034997 & S743000311 & 14:29:04.38 & $-$88:38:43.7 & 12.13 &  0.61 &  5.88 &  0.646577 &  LS & 785.6753 & RL & A   \\ 
n058442 &    \nd & S3YB000199 & 08:53:45.85 & $-$88:26:33.0 & 12.41 &  0.07 &  0.21 &  0.258295 &  LS & 785.2335 & PR &     \\ 
n058656 & 035468 & S3YA000613 & 08:08:46.28 & $-$88:00:02.0 & 13.86 &  0.17 &  0.09 &  0.822773 & BLS & 786.0050 & ED &     \\ 
n059543 & 036162 & S3YB000243 & 09:03:59.29 & $-$88:33:07.6 & 11.40 &  0.20 &  1.18 &  0.873857 &  LS & 785.2020 & ES & A   \\ 
n060041 & 036526 & S743000153 & 15:35:01.14 & $-$88:16:11.9 & 13.16 &  0.95 &  4.98 &    \nd    & \nd &   \nd    & LT &     \\ 
n060566 & 036939 & S743000200 & 15:25:11.74 & $-$88:23:14.9 & 12.82 &  0.05 &  0.05 &  7.376734 &  LS & 785.7479 & PR &     \\ 
n060667 & 037016 & S743000186 & 15:28:41.36 & $-$88:21:44.0 & 13.63 &  0.15 &  0.10 &  0.157962 &  LS & 785.3571 & PR &     \\ 
n060789 &    \nd & S741000025 & 15:59:17.54 & $-$88:00:42.5 & 14.23 &  0.26 &  0.03 &  6.853790 &  LS & 789.0467 & ED &     \\ 
n062144 & 038255 & S743000115 & 13:53:18.49 & $-$88:54:14.6 & 12.87 &  0.47 &  2.17 &  0.266903 &  LS & 785.2292 & EC & A   \\ 
n062519 & 038580 & S3Y8000346 & 10:50:11.96 & $-$88:59:54.1 & 11.78 &  0.05 &  0.21 &  5.745575 &  LS & 787.1325 & PR &     \\ 
n062640 & 038663 & S3YB000253 & 08:46:12.64 & $-$88:33:42.9 & 12.00 &  0.44 &  5.22 &  0.267127 &  LS & 785.2565 & EC & A   \\ 
n062696 &    \nd & S741000340 & 16:07:56.96 & $-$87:59:09.7 & 11.98 &  0.08 &  0.31 &  1.185502 &  LS & 785.2992 & PR &     \\ 
n062884 &    \nd & S740000057 & 12:30:49.10 & $-$89:02:45.7 & 11.94 &  0.03 &  0.06 & 25.125450 &  LS & 787.7145 & PR &     \\ 
n063743 & 039541 & S740000060 & 12:36:24.67 & $-$89:04:02.7 & 10.97 &  0.04 &  0.46 & 25.161758 &  LS &   \nd    & MP &     \\ 
n064731 & 040351 & S3YA000660 & 08:02:26.52 & $-$88:14:26.1 & 13.34 &  0.11 &  0.13 &  0.285606 &  LS & 785.4430 & PR &     \\ 
n065137 &    \nd & S74E000575 & 16:31:36.82 & $-$87:40:07.7 & 12.11 &  0.09 &  0.27 &  5.388951 &  LS & 790.4191 & ED &     \\ 
n066680 &    \nd & S3YL000291 & 07:14:42.32 & $-$87:22:25.0 & 11.37 &  0.07 &  0.49 &    \nd    & \nd &   \nd    & IR &     \\ 
n066932 &    \nd & S743000330 & 15:21:39.34 & $-$88:41:59.3 & 13.00 &  0.06 &  0.07 & 20.753202 &  LS & 785.2020 & PR &     \\ 
n068660 & 043618 & S3YB009826 & 08:40:28.89 & $-$88:47:00.4 & 13.72 &  0.33 &  0.26 & 13.024607 &  LS & 794.0587 & ES &     \\ 
n069025 & 043885 & S3YB000275 & 08:17:16.96 & $-$88:37:29.5 & 14.00 &  0.17 &  0.09 &  9.386448 &  LS & 785.9529 & PR &     \\ 
n070113 & 044751 & S740000101 & 13:01:29.04 & $-$89:13:47.0 & 13.21 &  0.11 &  0.14 &  6.422671 &  LS & 787.0902 & PR &     \\ 
n073013 & 047176 & S3YA000181 & 07:12:14.54 & $-$87:51:19.7 & 12.81 &  0.15 &  0.42 &  2.643748 &  LS & 786.7776 & PR &     \\ 
n076289 &    \nd & S3YA000320 & 07:07:52.74 & $-$88:02:11.4 & 12.94 &  0.10 &  0.13 &  0.789882 &  LS & 785.4483 & PR &     \\ 
n076976 & 050375 & S3YB013754 & 07:39:17.11 & $-$88:40:41.7 & 13.46 &  0.10 &  0.06 &  1.441150 &  LS & 785.6660 & PR &     \\ 
n077512 & 050773 & S74E000469 & 17:10:42.44 & $-$87:30:09.3 & 10.52 &  0.06 &  1.12 &  3.292027 &  LS &   \nd    & MP &     \\ 
n078642 &    \nd & S3Y8000239 & 08:51:51.07 & $-$89:15:21.3 & 12.27 &  0.05 &  0.09 &  3.481819 &  LS & 785.5641 & PR &     \\ 
n080090 & 052891 & S741000155 & 17:13:18.69 & $-$87:42:28.6 &  9.44 &  0.17 &  8.41 & 59.365516 &  LS &   \nd    & MP & A   \\ 
n080626 &    \nd & S743000493 & 16:37:09.46 & $-$88:43:24.2 & 11.29 &  0.04 &  0.23 & 35.753464 &  LS &   \nd    & MP &     \\ 
n080649 &    \nd & S3YA000502 & 07:01:38.41 & $-$88:17:02.9 &  9.60 &  0.05 &  1.52 & 23.466222 &  LS &   \nd    & MP &     \\ 
n080770 & 053446 & S3Y8000220 & 08:05:03.24 & $-$89:07:54.3 & 12.95 &  0.13 &  0.20 &  1.071988 &  LS & 786.1217 & PR &     \\ 
n080929 & 053570 & S743000460 & 16:44:10.40 & $-$88:38:19.9 & 11.02 &  0.03 &  0.26 & 19.967817 &  LS &   \nd    & MP &     \\ 
n081202 & 053783 & S3Y8000272 & 09:44:14.75 & $-$89:28:17.6 & 10.48 &  0.02 &  0.18 & 16.047868 &  LS &   \nd    & MP &     \\ 
n083000 &    \nd & S3Y8007194 & 09:23:04.69 & $-$89:29:38.8 & 13.77 &  0.23 &  0.25 &  0.124657 &  LS & 785.2675 & PR &     \\ 
n083359 & 055495 & S3Y8000078 & 07:43:54.49 & $-$89:07:37.3 & 12.52 &  0.39 &  1.45 &  0.797910 &  LS & 785.8882 & EC & A   \\ 
n084427 & 055854 & S3Y8000109 & 07:54:37.65 & $-$89:15:40.9 &  9.75 &  0.08 &  1.67 &    \nd    & \nd &   \nd    & IR &     \\ 
n085632 & 057344 & S741000539 & 17:13:42.52 & $-$88:24:52.5 & 11.09 &  0.09 &  1.60 & 20.106163 &  LS &   \nd    & MP &     \\ 
n086263 & 057775 & S3YA000492 & 06:40:47.15 & $-$88:15:21.3 & 11.73 &  0.39 &  4.71 &  0.438659 &  LS & 785.4511 & EC & A   \\ 
n088489 &    \nd & S3YA016263 & 06:31:32.66 & $-$88:11:38.1 & 12.81 &  0.09 &  0.13 &  0.240723 &  LS & 785.4360 & PR &     \\ 
n088653 & 059811 & S3YA000336 & 06:28:42.76 & $-$88:02:41.7 & 12.35 &  0.12 &  0.23 &  7.254001 &  LS & 787.3637 & ED &     \\ 
n090586 & 061353 & S740000342 & 17:15:45.51 & $-$89:00:42.8 & 10.78 &  0.02 &  0.18 &  0.022641 &  LS & 785.2130 & PR &     \\ 
n090919 & 061658 & S741000489 & 17:36:45.98 & $-$88:14:10.5 & 11.31 &  0.03 &  0.24 &  0.076166 &  LS & 785.2522 & PR &     \\ 
n091083 & 061740 & S740000322 & 17:23:25.05 & $-$88:53:37.3 &  9.99 &  0.07 &  2.20 & 34.479083 &  LS &   \nd    & MP &     \\ 
n091084 & 061783 & S740000308 & 17:25:37.17 & $-$88:49:50.3 & 13.83 &  0.35 &  0.35 &  0.188993 &  LS & 785.3176 & DS &     \\ 
n092211 & 062683 & S3Y8000165 & 07:46:18.25 & $-$89:40:00.7 & 10.27 &  0.04 &  0.75 & 15.857866 &  LS &   \nd    & MP &     \\ 
n094368 & 064380 & S3YA000248 & 06:10:28.12 & $-$87:53:32.9 & 14.20 &  0.24 &  0.03 &  2.307363 &  LS & 785.2020 & PR &     \\ 
n095083 & 064944 & S741000460 & 17:51:13.16 & $-$88:09:48.8 & 10.65 &  0.05 &  0.96 & 30.763139 &  LS &   \nd    & MP &     \\ 
n095145 & 065072 & S740000301 & 17:47:26.35 & $-$88:46:07.9 & 11.87 &  0.05 &  0.18 &  0.620534 &  LS & 785.6897 & PR &     \\ 
n096554 & 066196 & S740000469 & 17:05:16.14 & $-$89:51:43.8 &  9.67 &  0.04 &  1.06 & 26.623757 &  LS &   \nd    & MP &     \\ 
n097049 &    \nd & S740012990 & 17:55:19.53 & $-$89:07:43.2 & 14.97 &  0.56 &  0.14 &  0.348088 &  LS & 785.2424 & PR &     \\ 
n097333 & 066775 & S741000378 & 17:59:00.73 & $-$88:01:32.9 & 11.76 &  0.18 &  1.86 & 38.853950 &  LS &   \nd    & MP &     \\ 
n099251 & 068276 & SA9S000144 & 18:22:33.29 & $-$89:36:22.9 & 11.39 &  0.03 &  0.20 &  2.835357 &  LS & 786.8351 & PR &     \\ 
n099286 & 068308 & S0SG000353 & 05:50:34.20 & $-$89:06:46.2 & 12.94 &  0.08 &  0.10 &  0.798947 &  LS & 785.3570 & PR &     \\ 
n100083 & 068908 & SA9U000383 & 18:08:15.09 & $-$88:18:02.9 & 10.83 &  0.03 &  0.33 &  2.842765 &  LS &   \nd    & MP &     \\ 
n101873 &    \nd & S0SH000511 & 05:43:48.57 & $-$88:32:56.9 & 12.75 &  0.12 &  0.22 &  0.646457 &  LS & 785.3882 & PR &     \\ 
n102304 & 070680 & SA9S000413 & 22:05:02.55 & $-$89:52:06.7 & 13.05 &  0.21 &  0.56 &  1.988474 &  LS & 787.1308 & ES &     \\ 
n102641 & 070941 & S0SH000215 & 05:47:08.05 & $-$87:51:00.2 & 10.23 &  0.05 &  1.21 &  0.606546 &  LS & 785.4046 & GD &     \\ 
n103401 & 071571 & S0SH000333 & 05:43:19.93 & $-$88:04:04.3 & 10.65 &  0.32 &  6.76 &    \nd    & \nd &   \nd    & LT & A   \\ 
n104524 & 072350 & SA9V000050 & 18:30:57.87 & $-$88:43:17.5 &  9.86 &  0.04 &  0.57 &  9.925551 & BLS & 788.2407 & TR &     \\ 
n104943 & 072730 & SA9U000438 & 18:29:03.93 & $-$88:32:31.9 & 13.32 &  0.51 &  1.85 &  0.573044 &  LS & 785.7601 & RL & A   \\ 
n105244 &    \nd & S0SG000150 & 00:20:19.58 & $-$89:48:38.0 &  9.43 &  0.04 &  1.22 & 10.923423 &  LS &   \nd    & MP &     \\ 
n108372 &    \nd & SA9S000233 & 20:34:48.30 & $-$89:34:03.2 & 12.81 &  0.07 &  0.12 &  2.371811 &  LS & 785.4818 & PR &     \\ 
n109325 & 076135 & SA9S000115 & 19:53:21.15 & $-$89:22:46.5 & 10.37 &  0.33 &  8.00 &    \nd    & \nd &   \nd    & IR &     \\ 
n110028 & 076723 & SA9V000063 & 19:00:22.60 & $-$88:47:53.7 & 10.88 &  0.21 &  4.41 &    \nd    & \nd &   \nd    & IR &     \\ 
n110665 & 077190 & S0SH000448 & 05:15:49.62 & $-$88:17:51.5 & 10.21 &  0.04 &  0.92 & 11.462763 &  LS &   \nd    & MP &     \\ 
n111158 & 077508 & SA9S000404 & 22:53:06.46 & $-$89:38:40.9 & 14.07 &  0.36 &  0.09 &  0.133318 &  LS & 785.3185 & PR &     \\ 
n111246 & 077594 & SA9V000067 & 19:08:12.87 & $-$88:49:18.5 & 10.06 &  0.02 &  0.48 &  1.679604 &  LS & 785.5144 & PR &     \\ 
n112440 & 078549 & S0SH000321 & 05:15:49.03 & $-$88:04:23.5 & 13.12 &  0.16 &  0.42 &  5.389812 &  LS & 787.9077 & PR &     \\ 
n112694 & 078773 & SA9V000073 & 19:17:53.08 & $-$88:51:11.2 & 11.81 &  0.09 &  0.68 &  0.372068 &  LS & 785.2474 & EC &     \\ 
n113158 &    \nd & SA9S000043 & 19:33:21.70 & $-$89:00:22.9 &  8.67 &  0.17 &  2.29 &    \nd    & \nd &   \nd    & IR & A   \\ 
n113486 & 079397 & SA9S000068 & 19:48:57.10 & $-$89:07:14.3 & 10.86 &  0.02 &  0.21 &  4.842585 &  LS & 788.1947 & PR &     \\ 
n114115 &    \nd & S0SG000229 & 03:01:17.31 & $-$89:24:30.2 & 12.38 &  0.05 &  0.08 &  5.767268 & BLS & 785.6845 & TR &     \\ 
n115348 & 080934 & SA9U000331 & 18:58:35.51 & $-$88:12:55.2 & 11.72 &  0.24 &  2.83 & 62.185489 &  LS &   \nd    & MP &     \\ 
n116034 &    \nd & S0SH000139 & 05:12:14.90 & $-$87:44:42.1 &  8.43 &  0.13 &  8.84 &    \nd    & \nd &   \nd    & IR & A   \\ 
n116344 & 081723 & SA9U000071 & 18:48:23.75 & $-$87:43:16.4 & 11.49 &  0.07 &  0.67 &  0.841671 &  LS & 785.5572 & GD &     \\ 
n116380 & 081749 &        \nd & 19:11:53.61 & $-$88:27:14.9 & 10.31 &  0.03 &  0.54 & 88.385466 &  LS &   \nd    & MP &     \\ 
n116491 & 081845 & SA9V000074 & 19:38:27.80 & $-$88:50:55.9 & 12.90 &  0.10 &  0.09 &  6.759592 &  LS & 785.3202 & ED &     \\ 
n116918 & 082180 & S0SG000129 & 00:29:58.19 & $-$89:30:20.3 & 11.24 &  0.02 &  0.15 &  0.152825 &  LS & 785.3478 & DS &     \\ 
n117549 &    \nd & S0SG000213 & 02:55:59.73 & $-$89:17:48.2 & 12.48 &  0.06 &  0.11 &  0.721620 &  LS & 785.4964 & PR &     \\ 
n117763 &    \nd & SA9U000396 & 19:13:28.77 & $-$88:21:56.0 & 12.84 &  0.12 &  0.27 &  0.151852 &  LS & 785.2020 & DS &     \\ 
n118098 & 083110 & S0SG000226 & 02:11:02.80 & $-$89:22:50.3 & 10.27 &  0.03 &  0.43 &  6.623027 &  LS &   \nd    & MP &     \\ 
n118528 &    \nd & SA9S000384 & 23:35:11.99 & $-$89:27:51.8 & 14.78 &  1.23 &  0.86 &  0.465790 &  LS & 785.3995 & RL & G   \\ 
n118992 &    \nd & SA9S000200 & 21:24:11.87 & $-$89:17:56.5 & 14.69 &  0.52 &  0.15 &  0.233625 &  LS & 785.2898 & DS &     \\ 
n119729 & 084344 & SA9S000371 & 23:01:30.03 & $-$89:25:01.5 & 13.96 &  0.18 &  0.07 &  2.927708 & BLS & 786.0597 & PR &     \\ 
n121493 & 085719 & S0SG000178 & 03:08:19.12 & $-$89:06:32.2 & 12.20 &  0.06 &  0.18 & 17.205471 &  LS &   \nd    & MP &     \\ 
n122451 & 086480 & S0SJ000306 & 03:44:10.88 & $-$88:52:09.5 & 11.19 &  0.03 &  0.22 & 21.496218 &  LS &   \nd    & MP &     \\ 
n122836 &    \nd & SAA5000323 & 19:02:33.97 & $-$87:35:30.2 &  8.86 &  0.07 &  2.42 &    \nd    & \nd &   \nd    & IR &     \\ 
n123187 & 087084 & SA9S000168 & 20:57:31.47 & $-$89:03:50.3 & 12.36 &  0.33 &  0.76 &  1.857642 &  LS & 786.7030 & ED &     \\ 
n123522 &    \nd & SA9U000336 & 19:27:57.26 & $-$88:13:26.2 &  8.80 &  0.32 & 17.60 &    \nd    & \nd &   \nd    & IR & A   \\ 
n123706 & 087501 & S0SG000092 & 01:23:01.27 & $-$89:17:09.4 & 13.60 &  0.20 &  0.12 &  0.193489 &  LS & 785.3819 & DS &     \\ 
n123782 & 087548 & S0SH000485 & 04:20:11.85 & $-$88:25:03.5 & 12.63 &  0.12 &  0.61 &  0.197741 &  LS & 785.2566 & DS &     \\ 
n124517 & 088142 & S0SG000093 & 00:52:40.76 & $-$89:17:32.4 & 13.84 &  0.38 &  0.41 &  0.292944 &  LS & 785.3965 & EC &     \\ 
n126026 & 089391 & SA9U000387 & 19:43:40.77 & $-$88:19:20.7 & 13.49 &  0.08 &  0.04 &  0.136874 &  LS & 785.3534 & PR &     \\ 
n128266 &    \nd & S0SJ000248 & 03:37:18.58 & $-$88:38:51.4 &  9.47 &  0.04 &  0.95 &  7.957924 &  LS &   \nd    & MP &     \\ 
n130449 &    \nd & S0SG000064 & 00:24:32.27 & $-$89:08:48.5 & 12.69 &  0.08 &  0.16 &  4.314518 &  LS & 789.0505 & PR &     \\ 
n131494 & 093873 & SA9S000300 & 22:17:44.44 & $-$89:01:38.1 &  9.45 &  0.10 &  4.05 &    \nd    & \nd &   \nd    & IR &     \\ 
n132580 & 094793 & S0SU000407 & 04:36:31.02 & $-$87:28:03.3 & 10.97 &  0.09 &  1.50 &  0.617076 &  LS & 785.5155 & PR &     \\ 
n132830 &    \nd & S0SH031178 & 04:00:54.92 & $-$88:09:59.6 & 14.40 &  0.25 & -0.00 &  0.159051 & BLS & 785.2264 & PR &     \\ 
n133214 &    \nd & S0SJ000325 & 02:03:09.66 & $-$88:55:31.4 & 12.04 &  0.04 &  0.13 &  0.444790 &  LS & 785.3150 & PR &     \\ 
n134610 & 096404 & SAA5000420 & 19:43:05.07 & $-$87:46:52.4 & 14.13 &  0.81 &  1.38 &  0.581666 &  LS & 785.3416 & RL & A   \\ 
n134728 &    \nd & S0SJ014770 & 03:14:01.11 & $-$88:32:53.1 & 14.42 &  0.39 &  0.07 &  1.209346 &  LS & 785.8792 & ED &     \\ 
n135670 & 097230 & SA9U000295 & 20:02:18.84 & $-$88:02:50.0 & 12.08 &  0.10 &  0.21 &  9.548319 &  LS & 789.9965 & ED &     \\ 
n136750 & 098092 & S0SH000022 & 03:54:41.13 & $-$88:02:50.6 & 11.61 &  0.16 &  2.12 & 74.090207 &  LS &   \nd    & MP & A   \\ 
n137559 & 098719 & SAA5000417 & 19:50:26.13 & $-$87:44:50.7 & 12.86 &  0.25 &  1.07 &  0.416436 &  LS & 785.3327 & EC &     \\ 
n138555 & 099529 & S0SG000018 & 00:31:15.83 & $-$88:55:17.9 & 10.80 &  0.03 &  0.27 &  9.422044 &  LS &   \nd    & MP &     \\ 
n140799 &    \nd & S0SJ016907 & 02:29:57.18 & $-$88:34:35.3 & 14.32 &  0.22 & -0.01 &  2.852904 &  LS & 787.3254 & PR &     \\ 
n142023 &    \nd & S0SH031869 & 03:31:30.30 & $-$88:04:24.4 & 14.45 &  0.34 &  0.07 &  0.722102 &  LS & 785.7695 & ES &     \\ 
n142074 &    \nd & SA9V000177 & 21:19:45.21 & $-$88:28:17.7 & 13.39 &  0.16 &  0.21 & 43.638939 &  LS &   \nd    & MP &     \\ 
n142981 &    \nd & S0SJ000161 & 02:41:54.21 & $-$88:26:02.9 & 10.98 &  0.04 &  0.47 & 11.204593 &  LS &   \nd    & MP &     \\ 
n145245 &    \nd & S0SJ000052 & 00:56:07.00 & $-$88:42:17.9 & 11.36 &  0.04 &  0.25 &  7.994425 &  LS &   \nd    & MP &     \\ 
n145960 &    \nd & S0SJ000031 & 01:51:34.69 & $-$88:33:26.9 & 12.01 &  0.06 &  0.27 & 10.882430 &  LS &   \nd    & MP &     \\ 
n148233 & 107478 & SAA5000503 & 20:28:30.07 & $-$87:46:16.5 & 11.81 &  0.17 &  0.89 &  2.192580 &  LS & 786.0746 & ED &     \\ 
n148619 &    \nd & SA9V011956 & 22:28:21.41 & $-$88:31:41.0 & 14.34 &  0.22 &  0.01 &  3.817432 &  LS & 788.6046 & PR &     \\ 
n149414 &    \nd & S0SJ000002 & 03:00:33.53 & $-$88:02:59.2 & 10.13 &  0.86 & 29.50 &    \nd    & \nd &   \nd    & LT & A,G \\ 
n150808 &    \nd & SA9T000439 & 23:03:39.20 & $-$88:32:14.0 & 10.34 &  0.10 &  2.92 &    \nd    & \nd &   \nd    & IR & A   \\ 
n152437 & 110801 & S0SI000269 & 02:42:27.87 & $-$88:04:22.5 &  9.46 &  0.13 &  4.84 & 44.303305 &  LS &   \nd    & MP & A   \\ 
n153006 & 111298 & S0SI000438 & 02:12:56.05 & $-$88:13:52.5 & 10.03 &  0.04 &  1.11 & 12.116866 &  LS &   \nd    & MP &     \\ 
n156482 &    \nd & SA9T000422 & 23:57:27.17 & $-$88:24:54.5 & 12.44 &  0.06 &  0.11 &  6.198528 &  LS & 788.5551 & ED &     \\ 
n157627 &    \nd & S0SV000431 & 03:10:59.88 & $-$87:35:38.1 & 10.61 &  0.08 &  1.16 &    \nd    & \nd &   \nd    & IR &     \\ 
n159243 &    \nd & S0SI000372 & 00:00:50.89 & $-$88:19:41.9 &  9.44 &  0.04 &  1.12 &  0.114436 &  LS & 785.2310 & PR &     \\ 
n160137 & 117654 & S0ST000503 & 02:18:11.67 & $-$87:56:11.2 & 10.57 &  0.09 &  1.68 & 22.012562 &  LS &   \nd    & MP &     \\ 
n161425 & 118705 & S0SI033411 & 00:53:26.68 & $-$88:12:41.4 & 13.40 &  0.09 &  0.06 &  3.535402 & BLS & 786.7269 & TR &     \\ 
n162294 & 119488 & S0SI000329 & 00:08:43.43 & $-$88:13:48.4 & 10.07 &  0.11 &  3.59 &    \nd    & \nd &   \nd    & IR &     \\ 
n163148 & 120188 & S0SI000289 & 00:55:45.92 & $-$88:09:11.0 & 11.18 &  0.06 &  0.53 & 10.923423 &  LS &   \nd    & MP &     \\ 
n163888 &    \nd & SA9T000268 & 22:46:01.56 & $-$88:04:59.2 & 13.88 &  0.24 &  0.09 &  7.752665 &  LS & 787.3176 & ED &     \\ 
n164527 & 121369 & SA9T000310 & 23:15:35.46 & $-$88:07:33.8 & 10.78 &  0.05 &  0.75 & 24.945468 &  LS &   \nd    & MP &     \\ 
n167071 &    \nd & SAA6000049 & 21:52:26.36 & $-$87:44:20.6 & 10.59 &  0.05 &  0.82 & 10.560170 &  LS &   \nd    & MP &     \\ 
n167300 &    \nd & SAA7000327 & 21:12:08.31 & $-$87:24:25.1 &  9.17 &  0.08 &  2.21 &  0.166200 &  LS &   \nd    & MP &     \\ 
n167522 & 123934 & SA9T000282 & 23:52:30.36 & $-$88:03:49.5 & 13.94 &  0.14 &  0.04 &  0.122034 &  LS & 785.2471 & PR &     \\ 
n167944 &    \nd & S0SV012740 & 02:52:25.60 & $-$87:20:21.5 & 13.66 &  1.03 &  5.74 &    \nd    & \nd &   \nd    & LT &     \\ 
n168446 & 124666 & SAA6000034 & 21:47:16.29 & $-$87:39:06.6 & 12.84 &  0.62 &  3.88 &  0.458059 &  LS & 785.5620 & RL & A   \\ 
n171256 &    \nd & SA9T000182 & 23:16:45.97 & $-$87:54:16.6 & 11.95 &  0.07 &  0.29 &  8.754028 &  LS &   \nd    & MP &     \\ 
n172241 & 127850 & S0SI014387 & 00:00:52.46 & $-$87:54:27.3 & 11.38 &  2.91 & 61.18 &    \nd    & \nd &   \nd    & LT & G   \\ 
n173163 &    \nd & SAA6000447 & 22:12:01.66 & $-$87:37:04.2 &  9.57 &  0.07 &  2.75 &    \nd    & \nd &   \nd    & IR &     \\ 
n176396 &    \nd & SAA6000418 & 22:17:33.93 & $-$87:31:44.4 & 10.71 &  0.06 &  0.79 &    \nd    & \nd &   \nd    & IR &     \\ 
n177534 & 131919 & S0SI000101 & 00:01:16.84 & $-$87:44:02.9 & 11.99 &  0.09 &  0.24 &  9.458450 &  LS & 791.1664 & ED &     \\ 
n180169 & 133742 & SAA6000366 & 22:37:07.30 & $-$87:28:49.9 & 11.62 &  0.49 &  3.59 &  0.848393 &  LS & 785.5973 & EC & A   \\ 
n181332 &    \nd & SA9T000064 & 23:30:39.10 & $-$87:35:14.8 &  8.63 &  0.10 &  4.51 &    \nd    & \nd &   \nd    & IR & A   \\ 
n805954 & 005954 & S742000078 & 12:39:58.23 & $-$87:41:37.2 &  9.78 &  0.07 &  2.49 &    \nd    & \nd &   \nd    & IR &     \\ 
n863059 & 063059 & S3Y8000125 & 06:49:54.20 & $-$89:21:58.8 &  9.51 &  0.05 &  2.01 &    \nd    & \nd &   \nd    & IR &     \\ 
n897790 & 097790 & SA9V000415 & 22:23:40.80 & $-$88:53:42.9 &  9.77 &  0.03 &  0.78 &  0.521773 &  LS & 785.6966 & GD &     
\enddata
\tablecomments{[1]: from GSC2.3 except for \#n116380 (based on CSTAR master
  image) [2]: [LS], Lomb-Scargle method; [BLS], box-fitting algorithm.  [3]:
  Epoch of primary eclipse or minimum light in JD-2454500. [4]: Classes: [DS]:
  $\delta$~Scuti; [EC/ED/ES]: eclipsing binary (contact, detached,
  semi-detached); [GD]: $\gamma$ Doradus; [IR]: irregular; [LT]: long-term
  variation; [MP]: multi-periodic; [PR]: unclassified periodic; [RL]: RR
  Lyrae; [TR]: transit-like eclipse; [5]: [A]: in ASAS catalog; [G]: in GCVS.}
\end{deluxetable*}

\noindent{previous surveys. For example, our observations reach $i\sim15.3$~mag
  while the previous study of this area of the sky by ASAS reached a limiting
  magnitude of $V \sim 14.5$~mag (equivalent to $i\sim13-14.5$~mag depending on
  spectral type). We recovered 35 of the 46 previously-known variables in our
  field; 5 were saturated, 3 were strongly blended and 3 did not meet our
  minimum requirement for synoptic coverage.}

Our observations found significant variability or periodicity for 2.1\% of the
stars in the sample. This variable star fraction is in good agreement with the
expectation for a survey like ours with a photometric precision limit of $\sim
0.02$~mag \citep[see Fig.~9 of][]{Tonry2005}. Among the 44 variables selected
by the analysis of \S4.1, which did not discriminate by type of variability,
11\% are strictly periodic (eclipsing binaries, RR Lyraes, etc.) while an
additional 25\% exhibit multi-periodic behavior and the the remaining 64\%
have no significant periodicity.

Table~\ref{tab:vartypes} lists statistics for the different types of objects
we detected based on both searches (\S4.1 and \S4.2). The variable types are
consistent with expectations for a red-sensitive survey directed towards a
halo field. For example, nearly half of all variables exhibit
low-amplitude/multi-periodic or long-term/irregular pulsations typical of RGB
or AGB stars. 90\% of the variables of these types present in our sample have
$i<12$~mag, and according to the Besan\c{c}on model of the Galaxy
\citep{Robin2003} post-main sequence stars dominate the stellar population of
our field in that magnitude range. The regular pulsators that we have been
able to classify based on light curve shape or period are also post-main
sequence or Population II objects, such as RR Lyrae or $\delta$~Scuti
stars. In contrast, the eclipsing binaries and unclassified periodic variables
have a broader distribution in magnitude, as expected for a mix of evolved and
main sequence objects.

\begin{deluxetable}{lrr}
\tablewidth{0pt}
\tablecaption{Distribution of variable star types \label{tab:vartypes}}
\tablehead{\colhead{Variable Type} & \colhead{N} & \colhead{\%}}
\startdata
Multi-periodic         & 57 &  30.3 \\
Unclassified periodic  & 47 &  25.0 \\
Eclipsing binaries     & 35 &  18.6 \\
Irregular/long-term    & 28 &  14.9 \\
$\delta$~Sct           &  8 &   4.3 \\
RR Lyr                 &  7 &   3.7 \\
$\gamma$~Dor           &  3 &   1.6 \\
Transit-like           &  3 &   1.6 
\enddata
\tablewidth{0pt}
\tablecaption{Distribution of variable star types \label{tab:vartypes}}
\end{deluxetable}
 
\begin{figure}
\begin{center}
\includegraphics[height=0.49\textwidth]{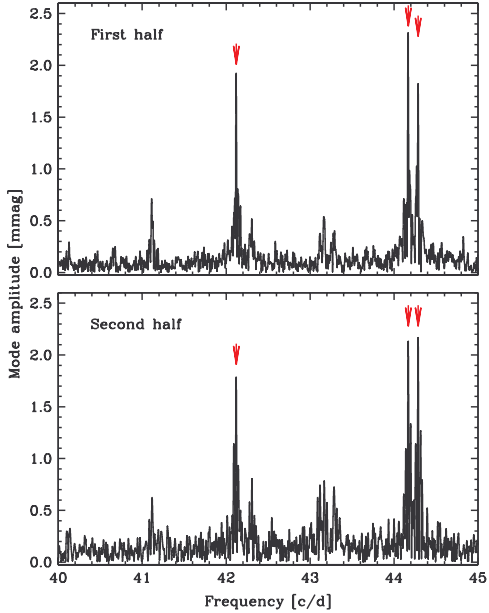}
\caption{Fourier spectrum of CSTAR \#n090586 derived using the Period04
  program, based on data obtained during the first (top) and second (bottom)
  half of the 2010 season. Three significant peaks (at $f_i = 44.288$,
  44.169, and 42.121 cycles d$^{-1}$) are detected with varying amplitudes,
  which exhibit significant changes ($4-11\sigma$) relative to the 2008 season.
  See text for details.\label{fig:fou}\vspace*{-12pt}}
\end{center}
\end{figure}

We combined the two years of CSTAR photometry (from the 2008 and 2010 Antarctic
winters) to search for changes in the observed properties of the variables. The
very fast periodic variable \#n090586 exhibited the same three significant
frequencies first seen in the 2008 observations \citep[see Fig.~17
of][]{Wang2011}, but the amplitude of each component exhibited significant
temporal variations relative to the that season (at the -3.8, 7.8 and
11.4$\sigma$ level, respectively). A comparison of the first and two halves of
the 2010 data (Fig.~\ref{fig:fou}) also shows a significant ($-3.9\sigma$)
variation for one of the frequencies. We plan further research on this object
using a fast-pulsating stellar model.

\begin{figure}
\begin{center}
\includegraphics[width=0.49\textwidth]{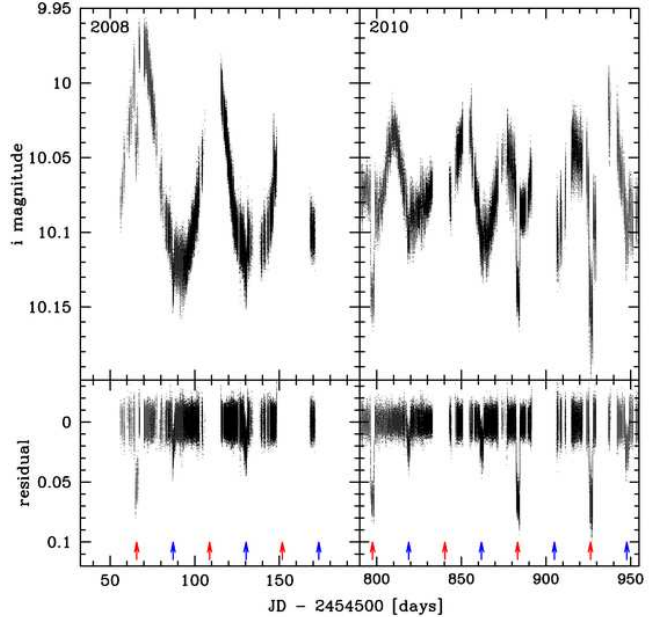}
\caption{Light curve of CSTAR\#n057725, showing the slow reduction in
  Cepheid-like pulsation amplitude during the 2008 season has been replaced by
  a more complex variability in 2010. Eclipse-like events seem to be present
  at two distinct phases indicated with red and blue arrows, respectively.\label{fig:cep}}
\end{center}
\end{figure}

The 2010 light curve of CSTAR\#n057725 ($P=43.2$~d), originally discovered by
the ASAS project \citep{Pojmanski2005} and classified as a fundamental-mode
$\delta$ Cephei variable, showed a dramatic change relative to 2008 as seen in
Fig.~\ref{fig:cep}. The Cepheid-like variability seen in the 2008 data
(top-left panel), which already showed indications of a change in amplitude,
decreased even further and developed a secondary peak in the 2010 data
(top-right panel). Furthermore, eclipse-like events can now be seen at two
distinct phases of the pulsation (bottom panels) separated by half of the period
with depths of $\sim 0.07$ and $\sim 0.03$~mag. The $V$-band ASAS light curve,
spanning more than a decade, also shows secular variations in pulsation
amplitude. The ASAS classification of this object is very unlikely; at
$i=10.1$~mag, a 43-day Cepheid would lie $\sim 17$~kpc away and nearly 8~kpc
above the Galactic plane, an extremely improbable location for an $\sim 11
M_{\odot}$ star. It is more likely that this object consists of a Population II
pulsator, such as an RV Tauri or the Galactic equivalent of an OGLE
small-amplitude variable red giant \citep{Soszynski2004} in a binary system. We
plan to undertake additional observations to investigate the nature of this system.

\section{Conclusions and Summary}

We have presented the analysis of high-cadence synoptic observations of 23
square degrees centered on the south celestial pole, conducted with the CSTAR\#3
telescope at Dome A during 183 days of the 2010 Antarctic winter season. 86\%
of all frames obtained at a sun elevation angle below $0^{\circ}$ yielded
useful data. We measured a median sky background of 19.8~mag $/\sq \arcsec$
over all moon phases and determined that the extinction is below 0.1~mag for
40\% of the time (0.4~mag for 70\% of the time). All these values are
consistent with the site statistics derived from the 2008 season, and
reinforce the promise of the Antarctic Plateau for future astronomical work.

We carried out time-series aperture photometry of 9,125 stars with
$i<15.3$~mag. We detected $\nvar$~variable stars, including $\nnew$~new objects
not included in our previous work thanks to a slightly larger field of view and
a deeper magnitude limit. This represents a $\sim 4\times$ increase in the
number of variable stars relative to previous surveys of the same region of the
sky. We plan follow-up multi-wavelength photometry and high-resolution
spectroscopy in the near future of the most interesting variables, such as
detached eclipsing binaries and the candidate transiting exoplanets.

Given the coarse pixel scale of our images and our use of aperture photometry,
the search for (non-periodic) variable stars was limited to objects with low
levels of crowding. This could mitigated by the use of difference imaging
techniques in future analyses. The apparent circular motion of stars in the
field of view, due to the fixed nature of the telescope and camera, prevented
the detection of periodic variability at frequencies close to one sidereal day
(and its harmonics) despite the absence of a day/night cycle and the long
duration of the observations. Future surveys conducted from Antarctica would
greatly benefit from instruments with improved angular resolution, a finer
pixel scale, and the ability to track. The recently-deployed AST3 telescopes
\citep{Cui2008SPIE-AST3} meets all these criteria and should greatly expand
our synoptic capabilities in the polar regions.

\ \par

Lingzhi Wang acknowledges support by: the BaiRen program of the Chinese Academy
of Sciences (034031001); the National Natural Science Foundation of China under
the Distinguished Young Scholar grant 10825313 and grants 11073005 \& 11303041;
the Ministry of Science and Technology National Basic Science Program (Project
973) under grant number 2012CB821804; the Excellent Doctoral Dissertation of
Beijing Normal University Engagement Fund; and a Young Researcher Grant of the
National Astronomical Observatories, Chinese Academy of Sciences.

Lucas Macri and Lifan Wang acknowledge support by the Department of Physics \&
Astronomy at Texas A\&M University through faculty startup funds and the
Mitchell-Heep-Munnerlyn Endowed Career Enhancement Professorship in Physics or
Astronomy.

\ \par

This work was supported by the Chinese PANDA International Polar Year project,
NSFC-CAS joint key program through grant number 10778706, CAS main direction
program through grant number KJCX2-YW-T08, and by the Chinese Polar Environment
Comprehensive Investigation \& Assessment Programmes (CHINARE). The authors
deeply appreciate the great efforts made by the 24-28th Dome A expedition teams
who provided invaluable assistance to the astronomers that set up and
maintained the CSTAR telescope and the PLATO system. PLATO was supported by the
Australian Research Council and the Australian Antarctic Division. Iridium
communications were provided by the US National Science Foundation and the US
Antarctic Program.

\ \par

{\it Facility:\ }\facility{Dome A: CSTAR}
 
\bibliography{wang}{}
\bibliographystyle{apj}

\end{document}